\documentclass[structabstract]{aa}
\usepackage{graphicx,natbib,amssymb}

\newcommand{\plotwd}{9.cm}
\newcommand{\cs}{\langle\sigma_A\upsilon\rangle}

\begin{document}
\title{Constraints on leptonically annihilating Dark Matter from reionization and extragalactic gamma background}
\titlerunning{Constraints on leptonically annihilating Dark Matter}
\author{Gert H\"utsi \inst{1,2}, Andi Hektor \inst{3}, Martti Raidal \inst{3}}
\institute{Department of Physics and Astronomy, University College London, London, WC1E 6BT\\
  \email{ghutsi@star.ucl.ac.uk} \and Tartu Observatory, T\~oravere 61602, Estonia \and National Institute of Chemical Physics and Biophysics, Tallinn 10143, Estonia\\
\email{andi.hektor@cern.ch, martti.raidal@cern.ch}}
\date{Received / Accepted}

\abstract
{The PAMELA,  Fermi and HESS experiments (PFH) have shown anomalous excesses in the cosmic positron and electron fluxes. A very exciting possibility is that those excesses are due to  annihilating dark matter (DM).}
{In this paper we calculate constraints on leptonically annihilating DM using observational data on diffuse extragalactic $\gamma$-ray background and measurements of the optical depth to the last-scattering surface, and compare those with the PFH favored region in the $m_{DM}-\cs$ plane.}
{Having specified the detailed form of the energy input with PYTHIA Monte Carlo tools we solve the radiative transfer equation which allows us to determine the amount of energy being absorbed by the cosmic medium and also the amount left over for the diffuse gamma background.}
{We find that the constraints from the optical depth measurements are able to rule out the PFH favored region fully for the $\tau^{-}+\tau^{+}$ annihilation channel and almost fully for the $\mu^{-}+\mu^{+}$ annihilation channel. It turns out that those constraints are quite robust with almost no dependence on low redshift clustering boost. The constraints from the $\gamma$-ray background are sensitive to the assumed halo concentration model and, for the power law model, rule out the PFH favored region for all leptonic annihilation channels. We also find that it is possible to have models that fully ionize the Universe at low redshifts. However, those models produce too large free electron fractions at $z\gtrsim 100$ and are in conflict with the optical depth measurements. Also, the magnitude of the annihilation cross-section in those cases is larger than suggested by the PFH data.}
{}
\maketitle

\section{Introduction}
The PAMELA and Fermi satellite experiments and the HESS atmospheric Cherenkov telescope have shown an anomalous excesses in the cosmic electron and positron spectra. PAMELA has observed a steep rise of positron fraction $e^+/(e^-+e^+)$ at energies
above 10 GeV with no significant excess in the cosmic antiproton flux \citep{Adriani:2008zq}.
 Fermi and HESS have measured an excess of high-energy $(e^-+e^+)$ flux 
 with a sharp cut-off of around 800 GeV \citep{2009Natur.458..607A,2009PhRvL.102r1101A,2009arXiv0906.2002H}. The ATIC and PPB-BETS balloon measurements indicate a similar excess \citep{2008Natur.456..362C,2008arXiv0809.0760T}. The most exciting explanation to those anomalies is that they might be caused by the annihilation of dark matter (DM) particles. However, the nature of those signals require the properties of DM to deviate strongly from the standard freeze-out predictions.  The DM annihilation cross-section $\cs$ has to be boosted some orders of magnitude over the standard freeze-out value of $\cs_{\rm std} \simeq 3 \times 10^{-26}\,{\rm cm}^3/s$ and the annihilations should favorably occur only through the leptonic channels.

Here we continue model independent studies of the DM annihilations  into Standard Model charged leptons, $DM + DM \to \ell^{-} + \ell^{+},$  $\ell=e,\,\mu,\,\tau,$ extending the analyses presented in  series of works, e.g. \citet{2009NuPhB.813....1C,2009arXiv0903.1852B,2009arXiv0904.3830C,2009arXiv0905.0480M,ArkaniHamed:2008qn}.
Because DM annihilations to charged particles necessarily induce $\gamma$-ray signal, the observed $\gamma$-ray
fluxes strongly constrain the DM annihilation  scenario as an explanation to the above mentioned cosmic ray anomalies.
Most of the  $\gamma$-ray constraints arise from the observations of Galactic center where the DM density is the highest.
Those analyses take into account both the primary  photons produced by  final state radiation and decays of 
the DM annihilation products $\ell$ \citep{Bertone:2008xr,2009PhRvD..79d3507B} as well as the secondary 
inverse Compton (IC) scattering photon flux produced by them \citep{2009arXiv0904.3830C,2009arXiv0905.0480M}.

In this study we focus on the possible effects of photons produced in extragalactic DM annihilations all over the Universe today and in the past. 
We are particularly interested in finding out if the extragalactic photon signals are able to rule out the PAMELA, Fermi, and HESS (PFH) favored regions on the $m_{DM}-\cs$ plane as given in \citet{2009arXiv0905.0480M}. For this purpose we look at two
extragalactic signals: (i) the diffuse $\gamma$-ray background including both the primary photons as well as the secondary
photon flux produced in IC scattering of electrons off the Cosmic Microwave Background (CMB) photons, (ii) the Thompson scattering optical depth of the  CMB photons caused by the free electrons between us and the last scattering surface~\footnote{In the following the observational constraint derived from the Thompson scattering optical depth is called the ``reionization constraint''.}.
While the previous works on those topics take into account only the direct photons produced in DM annihilations, we show that
in fact the diffuse IC photon spectrum gives the strongest constraint.
We stress the complementarity between those observables:  the more energy ends up in extragalactic gamma background the less energy is available for ionizing/heating the gas, and vice verse. 
The spectrum of diffuse $\gamma$-ray background has been measured by \citet{1998ApJ...494..523S} using data from the EGRET space experiment and the optical depth to decoupling has been derived from the CMB measurements by the WMAP satellite \citep{2009ApJS..180..306D}. New more precise data on $\gamma$-ray background is expected to be published by Fermi satellite this year.

The effect of  annihilating DM on reionization and recombination has been previously investigated by a few authors, e.g. \citet{2005PhRvD..72b3508P,2006MNRAS.369.1719M,2006PhRvD..74j3519Z,2008PhRvD..78j3524N,2009arXiv0904.1210B,2009arXiv0905.0003G,2009arXiv0906.1197S}. Prior to the calculations done in \citet{2008PhRvD..78j3524N,2009arXiv0904.1210B} all the authors have only included the homogeneous DM component, i.e. have neglected the clustering effect. Despite the inclusion of the clustering there are some problems with the formalism of \citet{2008PhRvD..78j3524N,2009arXiv0904.1210B}: the treatment for the evolution of the matter temperature is simplistic. Also, in \citet{2008PhRvD..78j3524N}, the optical depth \footnote{The argument of $\exp$ in their Eq. (17).} for the $\gamma$-rays is calculated incorrectly. The original work by \citet{2005PhRvD..72b3508P}, and in particular, the extended newer version by \citet{2009arXiv0906.1197S} uses more accurate formalism, unfortunately the authors have not included the effect of DM clustering. 

During completion of our study couple of papers appeared calculating the diffuse $\gamma$-ray background due to the annihilating DM, e.g. \citet{2009arXiv0904.3626K,2009arXiv0906.0001P,2009arXiv0906.2251B}. \citet{2009arXiv0904.3626K} have neglected the $\gamma$-rays from the inverse Compton scattering of the final state $e^-/e^+$, which gives even stronger bound than the primary $\gamma$-rays. \citet{2009arXiv0906.0001P} have used different formalism for the absorption of $\gamma$-rays considering the numerical approximation from \citet{2004PhRvD..70d3502C}. In this paper we include the relevant absorption processes directly (see the discussions below).

Our paper is organized as follows: In Section 2 we describe the source term, i.e. the energy input from the DM annihilation. In Section 3 we solve the radiative transfer equation and present a few example $\gamma$-ray spectra along with some results for the evolution of the ionization fraction and matter temperature. Section 4 presents our main results and our summary is given in Section 5.

\section{Energy input from DM annihilation}
\begin{figure}
\centering
\includegraphics[width=\plotwd]{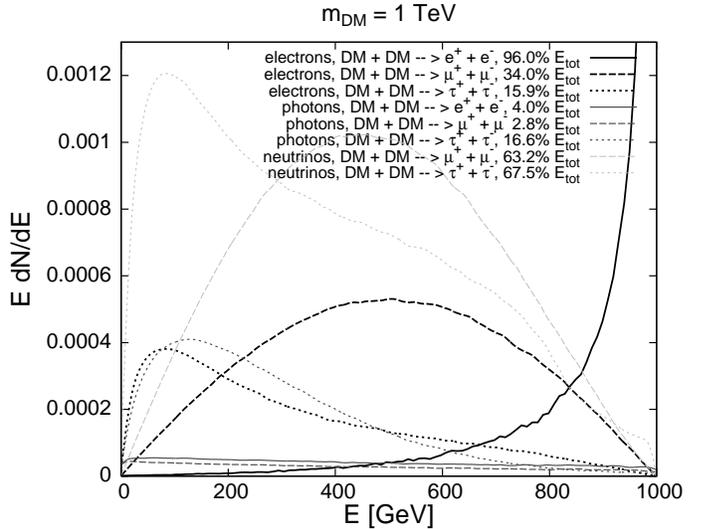}
\caption{The final state distributions of electrons/positrons, photons, and neutrinos/antineutrinos for the annihilating DM with $m_{DM}=1$ TeV.}
\label{fig1}
\end{figure}

In our study we consider the DM annihilation to Standard Model charged leptons following the leptonic annihilation scenario presented in \citet{2009NuPhB.813....1C}. The final state distributions of electrons/positrons, photons, and neutrinos/antineutrinos from the primary two-body states produced in DM annihilations are computed using MonteCarlo code PYTHIA \citep{2008CoPhC.178..852S}. The example final particle distributions for all three annihilation channels assuming the input DM particle with mass $m_{DM}=1$ TeV are shown in Fig. \ref{fig1}. The partition of energy between final particles is shown in Table \ref{tab1}. The released electrons/positrons immediately interact with the CMB photons and upscatter those to high energies via the inverse Compton mechanism. Thus, in addition to the prompt photons released directly in the annihilation process, our photon spectrum contains the other part: the upscattered inverse Compton photons.

\begin{table}
\caption{Energy partition between electrons/positrons, photons and neutrinos/antineutrinos for all three annihilation channels assuming $m_{DM}=100$ GeV - $10$ TeV.}
\label{tab1}
\begin{tabular}{c|c|c|c}
channel & electrons & photons & neutrinos \\
\hline
\hline
$e^{-}+e^{+}$ & $\sim 96-97\,\%$ & $\sim 3-4\,\%$ & $0\,\%$\\
\hline
$\mu^{-}+\mu^{+}$ & $\sim 34\,\%$ & $\sim 2-3\,\%$ & $\sim 63-64\,\%$\\
\hline
$\tau^{-}+\tau^{+}$ & $\sim 16\,\%$ & $\sim 16-17\,\%$ & $\sim 67-68\,\%$
\end{tabular}
\end{table}

Having an annihilating DM particle with mass $m_{DM}$ and with a thermally averaged cross-section $\cs$ the emitted power per volume and per steradian, i.e. the emission coefficient can be given as
\begin{equation}\label{eq01}
\bar{\jmath}(z)=\frac{1}{4\pi}\frac{\cs}{2m_{DM}}\bar{\rho}_{DM,0}^2(1+z)^6\,.
\end{equation}
It is the emission coefficient for the homogeneous distribution of DM. To include the effect of DM clustering $\bar{\rho}_{DM}^2(z)$ should be replaced by $\langle\rho_{DM}^2(z)\rangle\equiv \bar{\rho}_{DM}^2(z)\langle(1+\delta(z))^2\rangle\equiv B(z)\bar{\rho}_{DM}^2(z)$. Here $\delta\equiv\rho/\bar{\rho}-1$ is the density contrast and we have also defined a halo boost factor $B(z)\equiv\langle(1+\delta(z))^2\rangle=1+\langle\delta^2(z)\rangle$. To calculate the boost factor we use the halo model which approximates the matter distribution in the Universe as a superposition of DM halos (for a comprehensive review on halo model see \citet{2002PhR...372....1C}). Within this model $B(z)$ can be given as
\begin{equation}\label{eq02}
B(z)=1+\frac{\Delta_c}{3\bar{\rho}_{m,0}}\int\limits_{M_{\rm min}}^{\infty}{\rm d}M\frac{{\rm d}n}{{\rm d}M}(M,z)f\left[c(M,z)\right]\,,
\end{equation} 
where $\bar{\rho}_{m,0}$ is the matter density at $z=0$, $\Delta_c=200$ is the overdensity at which the halos are defined, $M_{\rm min}$ is the minimum halo mass, $\frac{{\rm d}n}{{\rm d}M}(M,z)$ is the halo mass function, $c(M,z)$ represents the halo concentration parameter and the function $f(c)$ for the halos with the NFW density profile \citep{1997ApJ...490..493N} is given as
\begin{equation}
f(c)=\frac{c^3}{3}\left[1-\frac{1}{(1+c)^3}\right]\left[\log(1+c)-\frac{c}{1+c}\right]^{-2}\,.
\end{equation}

\begin{figure}
\centering
\includegraphics[width=\plotwd]{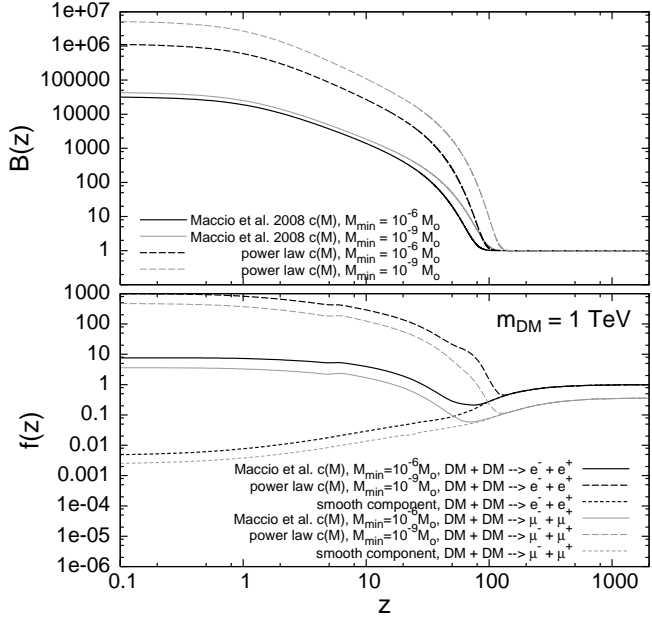}
\caption{(Upper panel) Boost factors $B(z)$ as defined in Eq. (\ref{eq02})) for various concentration models. (Lower panel) $f$-parameter as given by Eq. (\ref{eq10}), i.e. the ratio of the total energy deposition and local ``smooth'' energy injection rates for two different concentration models and two leptonic annihilation channels, assuming the annihilating DM particle with mass $m_{DM}=1$TeV. The short dashed lines show the ratio $\bar{\epsilon}/(4\pi \bar{\jmath})$ where for calculating $\bar{\epsilon}$ one takes $B=1$ in Eq. (\ref{eq06}). The results for the annihilation channel $DM+DM\rightarrow \tau^{-}+\tau^{+}$ are very similar to the $\mu^{-}+\mu^{+}$ case and for clarity are not shown here.}
\label{fig2}
\end{figure}

For the mass function we use the standard form given by \citet{1999MNRAS.308..119S}, for the concentration parameter we use two models: (i) the model by \citet{2008MNRAS.391.1940M}, (ii) the power law model where $c\propto M^{-0.1}$, which gives a good fit within the mass range resolved by the simulations. In our calculations we assume two values for the lower halo mass: $10^{-6}$ and $10^{-9}$ $M_{\odot}$ (for the motivation of those values see e.g. \citet{2009arXiv0903.0189B,2009JCAP...06..014M}). The resulting four models for the boost factor are given in the upper panel of Fig. \ref{fig2}. We see that the boost factor shoots up at $z\sim100$, which corresponds to the redshift where the DM halos with the assumed lowest masses start to form. As expected, the curves for the $10^{-9}$ $M_{\odot}$ models start to deviate slightly earlier. It is clear that currently the biggest uncertainty influencing the magnitude of the boost factor is related to the way one chooses to extend the behavior of the concentration parameter below the mass scales directly probed by the simulations. \footnote{For similar comments on uncertainties related to the mass dependence of the concentration parameter see e.g. \citet{2009JCAP...06..014M}.} We see that at low redshifts the variation between the models reaches two orders of magnitude. On top of that there is an extra uncertainty related to the profiles of the DM halos. However, this is expected to be of smaller magnitude, and mainly due to those reasons, we have decided to use only NFW profiles throughout this paper.

Using the boost factor we can write the total emission coefficient as
\begin{equation}\label{eq04}
j(z)=B(z)\bar{\jmath}(z)\,.
\end{equation}
We also define the corresponding spectral quantity
\begin{eqnarray}\label{eq05}
j_{\nu}(z)&=&j(z)f^z(\nu)\,,\\
\int f^z(\nu){\rm d}\nu&\equiv&1\,,\nonumber
\end{eqnarray} 
where the frequency distribution function $f^z(\nu)=f^{z}_{\rm IC}(\nu) + f^{z}_{\rm prompt}(\nu)$ consists of two parts: (i) Inverse Compton part which is produced by the energetic electrons/positrons interacting with the CMB. This is calculated following the formalism in \citet{2009arXiv0904.3830C}; (ii) Prompt photon part -- these are the photons released directly from the annihilation event.  

For the inverse Compton scattering $f^{z}_{\rm IC}(\nu)=\frac{1}{1+z}f^{z=0}_{\rm IC}(\frac{\nu}{1+z})$ and for the prompt emission $f^{z}_{\rm prompt}(\nu)=f^{z=0}_{\rm prompt}(\nu)$.

\section{Radiative transfer. Extragalactic gamma background. Reionization}
In the expanding Universe the formal solution of the radiative transfer equation for the intensity can be written in the form
\begin{eqnarray}\label{eq06}
I_{\nu}(z)&=&c\int\limits_{z}^{\infty}{\rm d}z^{'}\frac{1}{H(z^{'})(1+z^{'})}\left(\frac{1+z}{1+z^{'}}\right)^3j_{\nu^{'}}(z^{'})\cdot\\
&\cdot&\exp\left[-\tau_{\nu}(z,z^{'})\right]=\nonumber\\
&=&\frac{c \bar{\jmath}(z)}{(1+z)^3}\cdot\int\limits_{z}^{\infty}{\rm d}z^{'}\frac{(1+z^{'})^2}{H(z^{'})}B(z^{'})\cdot\nonumber\\
&\cdot&\left[f^{z^{'}}_{\rm IC}(\nu^{'})+f^{z^{'}}_{\rm prompt}(\nu^{'})\right]\exp\left[-\tau_{\nu}(z,z^{'})\right]\,.\nonumber
\end{eqnarray}
Here $j_{\nu}(z)$ is the emission coefficient given by Eqs. (\ref{eq01}), (\ref{eq04}), (\ref{eq05}), $H(z)$ is the Hubble parameter, factor $\propto (1+z)^{-3}$ accounts for the cosmological dimming of the intensity, and the optical depth $\tau_{\nu}(z,z^{'})$ is given as
\begin{equation}\label{eq07}
\tau_{\nu}(z,z^{'})=c\int\limits_{z}^{z^{'}}{\rm d}z^{''}\frac{\alpha_{\nu^{''}}(z^{''})}{H(z^{''})(1+z^{''})}\,.
\end{equation}
Here $\nu^{''}=\frac{1+z^{''}}{1+z}\nu$, and $\alpha_{\nu}(z)$ is the absorption coefficient. For $\alpha_{\nu}(z)$ we follow the formalism given in \citet{1989ApJ...344..551Z}. The following processes are included in our calculations:
\begin{itemize}
\item photoionization
\item Compton scattering
\item photon-matter pair production
\item photon-photon scattering
\item photon-photon pair production
\end{itemize}

\begin{figure}
\centering
\includegraphics[width=\plotwd]{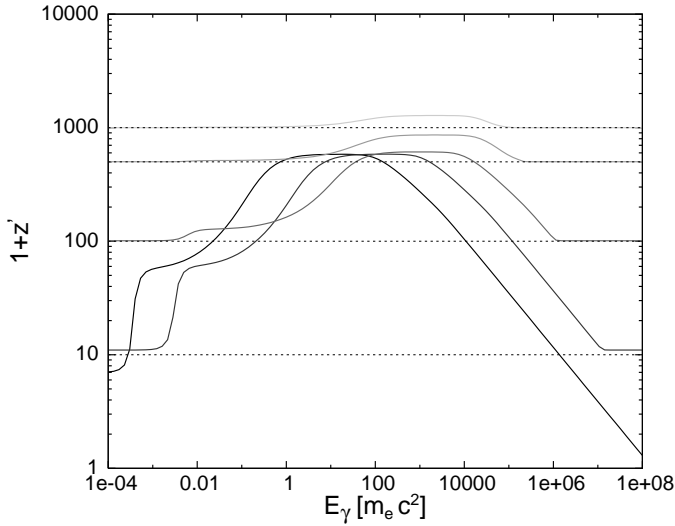}
\caption{The redshift $z^{'}$ where the optical depth for photons reaches unity (i.e., $\tau_{\nu}(z,z^{'})=1$ in Eq.(\ref{eq07})) for several ``observer's redshifts'': $z=0,10,100,500,1000$. Here the energy plotted is the photon energy at redshift $z$.}
\label{fig3}
\end{figure}

In Fig. \ref{fig3} we plot the redshift $z^{'}$ where the optical depth for photons reaches unity (i.e., $\tau_{\nu}(z,z^{'})=1$) for several ``observer's redshifts'': $z=0,10,100,500,1000$. Here the energy plotted is the photon energy at redshift $z$. The lowest curve is the analog of the curve plotted in Fig. 2 of \citet{1989ApJ...344..551Z}.

Having a method to calculate the intensity of the photon field sourced by the annihilating DM at each redshift we can go on and express the energy deposition rate in the cosmic medium as
\begin{equation}
\epsilon_{\nu}(z)=4\pi \alpha_{\nu}(z)I_{\nu}(z)\,.
\end{equation}
To estimate the fraction of the deposited energy ending up in ionizing and heating the medium we use the approximation motivated by the original work of \citet{1985ApJ...298..268S} and used in several subsequent papers \citep{2004PhRvD..70d3502C,2005PhRvD..72b3508P,2006MNRAS.369.1719M,2006PhRvD..74j3519Z,2008PhRvD..78j3524N}: $\sim (1-x_e)/3$ goes into ionization and $\sim (1 + 2x_e)/3$ into heating. Here $x_e$ is the ionization fraction. To calculate the evolution of the ionization fraction and matter temperature we modify the recombination code RECFAST \citep{1999ApJ...523L...1S} following the description in \citet{2005PhRvD..72b3508P}. For their modification we need the total energy deposition rate per hydrogen particle which is given as
\begin{equation}
\epsilon_H(z)=\frac{\epsilon(z)\equiv\int\limits_{0}^{\infty}\epsilon_{\nu}(z){\rm d}\nu}{n_{H,0}(1+z)^3}\,.
\end{equation}
To go beyond their ``on the spot'' approximation the $f$-parameter in their Eq. (5) should be replaced by the following function of $z$ (see also \citet{2009arXiv0906.1197S})     
\begin{equation}\label{eq10}
f(z)=\frac{\epsilon(z)}{4\pi \bar{\jmath}(z)}\,,
\end{equation}
which gives us the ratio of the total energy deposition and local ``smooth'' energy injection rates. The function $f(z)$ is plotted on the lower panel of Fig. \ref{fig2} for two different concentration models, two leptonic annihilation channels ($DM+DM\rightarrow e^{-}+e^{+}$, $DM+DM\rightarrow \mu^{-}+\mu^{+}$), assuming the annihilating DM particle with mass $m_{DM}=1$TeV. The upturn of the curves at redshift $z\sim100$ mimics the similar trend seen in the upper panel and is caused by the onset of the structure formation. The short dashed lines show the ratio $\bar{\epsilon}/(4\pi \bar{\jmath})$ where for calculating $\bar{\epsilon}$ one takes $B=1$ in Eq. (\ref{eq06}). These last two curves can be directly compared with the results presented in Fig. 4 of \citet{2009arXiv0906.1197S}. For clarity we have not shown the results for the annihilation channel $DM+DM\rightarrow \tau^{-}+\tau^{+}$ as those are very similar to the $\mu^{-}+\mu^{+}$ case. At large redshifts, where the Universe gets optically thick to photons, and where the structure boost $B(z)=1$, one expects $f(z)$ to asymptotically approach unity. This is indeed what we see in the case of $e^{-}+e^{+}$ annihilation channel. However, the asymptotic $f(z)$ values for the $\mu^{-}+\mu^{+}$ and $\tau^{-}+\tau^{+}$ cases are smaller since a large fraction of energy is carried away by neutrinos (see Table \ref{tab1}).

\begin{figure}
\centering
\includegraphics[width=\plotwd]{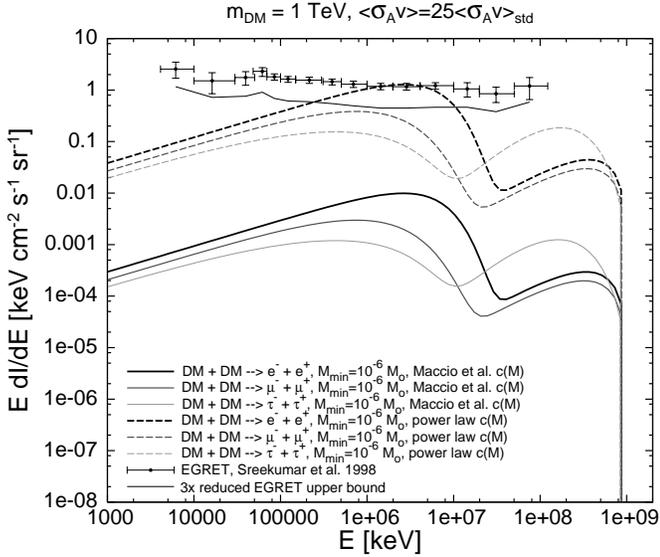}
\caption{Example $\gamma$-ray spectra at redshift $z=0$ assuming the annihilating DM particle with mass $m_{DM}=1$TeV. Here the two most extreme concentration models of Fig. \ref{fig2} have been used and the results are given for all three leptonic annihilation channels. The thermally averaged annihilation cross-section has been set to $25$ times its standard value of $\sim 3\times10^{-26}{\rm cm}^2$. The points with errorbars correspond to the EGRET measurements of the extragalactic gamma background as given in \citet{1998ApJ...494..523S}. The solid horizontal line, which is used in the following as an upper bound for the level of diffuse $\gamma$-ray background, represents the EGRET measurements reduced by a factor of three.}
\label{fig4}
\end{figure}

\begin{figure}
\centering
\includegraphics[width=\plotwd]{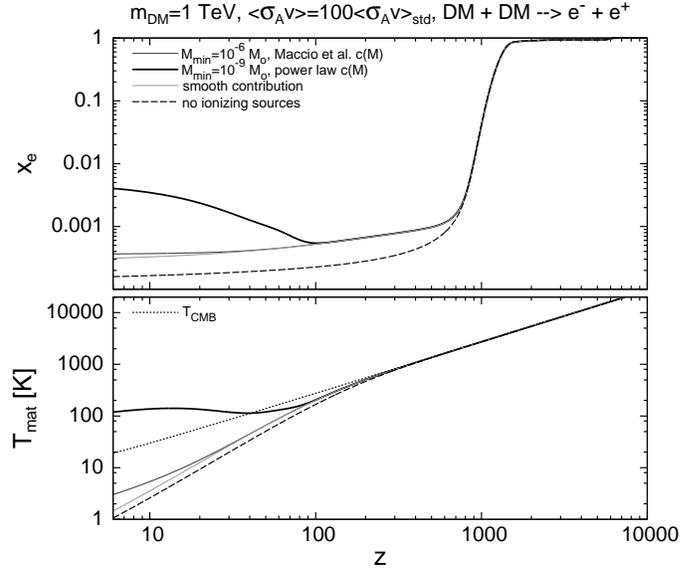}
\caption{The evolution of the ionization fraction (upper panel) and matter temperature (lower panel) for the model with the annihilating DM particle with mass $m_{DM}=1$TeV and with a thermally averaged cross-section $\cs=3\times10^{-24}{\rm cm}^2$. The two most extreme concentration models of Fig. \ref{fig2} have been used and for clarity only the results for the $e^{-}+e^{+}$ channel are shown. The lowest solid line plots the case where the structure boost has been neglected, i.e. $B(z)=1$ and the dashed line corresponds to the standard ``concordance'' cosmology without an annihilating DM. In the lower panel the dotted line shows the evolution of the CMB temperature.}
\label{fig5}
\end{figure}

In Fig. \ref{fig4} we show some example $\gamma$-ray spectra at redshift $z=0$ calculated with the above formalism for the annihilating DM particle with mass $m_{DM}=1$TeV. Here the two most extreme concentration models of Fig. \ref{fig2} have been used and the results are given for all three leptonic annihilation channels considered in this paper. The thermally averaged annihilation cross-section has been set to $25$ times its standard value of $\sim 3\times10^{-26}{\rm cm}^2$. The low energy bump of the characteristic double-bumped spectrum is due to the inverse Compton process whereas the high energy bump is the prompt photon contribution. The points with errorbars correspond to the EGRET measurements of the extragalactic gamma background as given in \citet{1998ApJ...494..523S}. The solid horizontal line, which is used in the following as an upper bound for the level of diffuse $\gamma$-ray background, represents the EGRET measurements reduced by a factor of three to approximately account for the following: (i) dependent on the modeling details, it is predicted that from $25\%$ up to $100\%$ of the diffuse extragalactic $\gamma$-ray background is due to unresolved AGNs \citep{1996ApJ...464..600S,1998ApJ...496..752C,1999APh....11..213M,2000MNRAS.312..177M}, (ii) the improved model for the Galactic contribution further suppresses the measurements \citep{2004ApJ...613..956S}, (iii) there is an additional component due to the DM annihilation produced in our own Galactic halo (see the relevant calculation by \citet{2009arXiv0904.3626K}).\footnote{If the Reader has her/his own favored value for the possible reduction factor the final results in Fig. \ref{fig6} are easily scalable to accommodate that, since the DM annihilation signal is simply proportional to $\cs$.}

Note that due to relatively small fraction of energy released in the form of prompt photons $\sim 2-4 \%$, except for the $\tau^{-}+\tau^{+}$ channel where the prompt photon part reaches up to $\sim 17 \%$, the inverse Compton part of the spectrum has a higher amplitude than the prompt portion. As the observational bounds are represented by a roughly horizontal line it is clear that in the case of $e^{-}+e^{+}$ and $\mu^{-}+\mu^{+}$ channels the constraint we obtain on $\cs$ arises from the inverse Compton portion only. However, for the $\tau^{-}+\tau^{+}$ channel and for relatively low $m_{DM}$, when the prompt bump of the spectrum reaches the energy range probed by EGRET, the constraint on $\cs$ is provided by the amplitude of the prompt portion, instead.

In Fig. \ref{fig5} we show an example of the evolution of the ionization fraction (upper panel) and matter temperature (lower panel) for the model with the annihilating DM particle with mass $m_{DM}=1$TeV and with a thermally averaged cross-section $\cs=3\times10^{-24}{\rm cm}^2$. The two most extreme concentration models of Fig. \ref{fig2} have been used and for clarity only the results for the $e^{-}+e^{+}$ channel are shown. The lowest solid line plots the case where the structure boost has been neglected, i.e. $B(z)=1$ and the dashed line corresponds to the standard ``concordance'' cosmology without an annihilating DM. We have shown the evolution only down to redshift $z=6$ beyond which there is a clear observational evidence that the Universe gets fully ionized \citep{2006ARA&A..44..415F}. In the lower panel the dotted line shows the evolution of the CMB temperature. Note that at $z\sim 100$ the power law concentration model with $M_{\rm min}=10^{-9}M_{\odot}$ clearly starts to deviate from the smooth model without the clustering boost, resulting in an order of magnitude higher ionization fraction at low redshifts. However, if one calculates the optical depth to the last scattering then all of the models with DM annihilation plotted in Fig. \ref{fig5} give very similar results, i.e. clustering boost at low $z$ has minimal effect. In \citet{2009arXiv0904.1210B} the authors discuss models where the DM annihilation is able to fully reionize the Universe at low redshifts. In our calculations we also find that, indeed, it is possible to fully reionize the Universe, however, all of those models give too large amplitudes for the ionization fraction plateau after $z\sim 100$ and thus violate the CMB measurements. 

\section{Results}

\begin{figure}
\centering
\includegraphics[width=\plotwd]{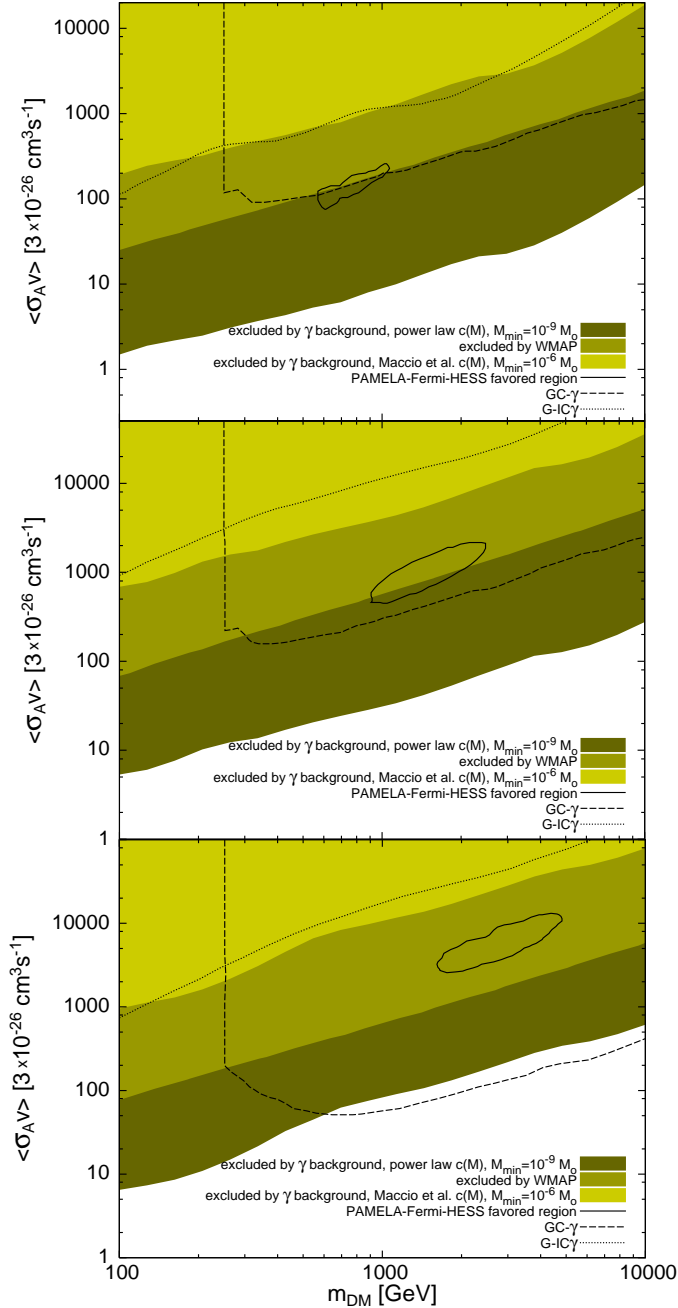}
\caption{Constraints from reionization and extragalactic gamma background for all three leptonic annihilation channels. The upper, middle, and lower panel correspond to $e^{-}+e^{+}$, $\mu^{-}+\mu^{+}$, and $\tau^{-}+\tau^{+}$ channels, respectively. The solid line shows the PAMELA, Fermi, and HESS favored region as given in \citet{2009arXiv0905.0480M}. We have also plotted some Galactic constraints taken from \citet{2009arXiv0905.0480M}: dashed lines show the bounds from the HESS measurements of the Galactic Center $\gamma$-rays, dotted lines are the bounds from FERMI measurements of $\gamma$-rays below $10$ GeV. (See the main text for further details.)}
\label{fig6}
\end{figure}

In this section we are going to confront the leptonically annihilating DM models with the observational data. We are going to use: (i) measurements of the extragalactic $\gamma$-ray background by EGRET \citep{1998ApJ...494..523S}, (ii) measurement of the optical depth to the last scattering surface by WMAP \citep{2009ApJS..180..306D}. As described in the previous section, to account for the better Galactic subtraction due to \citet{2004ApJ...613..956S}, for the annihilation signal from our own Galaxy, and for the contribution of AGNs to the $\gamma$-ray background, we use the upper bound of \citet{1998ApJ...494..523S} reduced by a factor of three as our upper limit for the possible signal rising from the annihilating DM. This upper bound is shown as a solid horizontal line in Fig. \ref{fig4}.

To calculate the constraint from reionization we use the WMAP measurement for the optical depth: $\tau=0.087 \pm 0.017$. Using the spectra of distant quasars we know that most of the hydrogen in the Universe is ionized below redshift $z\simeq6$ \citep{2006ARA&A..44..415F}. As was also done in \citet{2008PhRvD..78j3524N,2009arXiv0904.1210B} we assume that helium is singly ionized below $z=6$ and doubly ionized below $z=3$. This assumption implies that the contribution to the WMAP optical depth from redshifts $z>6$ is $\tau(z>6)=0.047\pm0.017$. In the following we use the $1-\sigma$ upper bound $\tau(z>6)=0.064$ for deriving our constraints. The constraints we find on a $m_{DM}-\cs$ plane for all three leptonic annihilation channels are given in Fig. \ref{fig6}. Here the upper, middle, and lower panel correspond to $e^{-}+e^{+}$, $\mu^{-}+\mu^{+}$, and $\tau^{-}+\tau^{+}$ channels, respectively. As it turns out, our constraints from WMAP $\tau$ measurements are basically independent of the low redshift clustering boost, implying that already the high redshift (i.e. $z\gtrsim 100$) energy injection from the unclustered component is able to account for the measured level of $\tau(z>6)$.

We also note that, indeed, as pointed out in \citet{2009arXiv0904.1210B} one can have models where the annihilating DM is able to fully reionize the Universe at low redshifts. However, we find that those models lead to the strong violation of the WMAP optical depth measurement due to the vastly increased level of the ionization fraction plateau beyond redshifts $z\sim 100$.

In Fig. \ref{fig6} we have also plotted regions favored by the PFH data as given in \citet{2009arXiv0905.0480M}. We see that for the power law concentration model with $M_{\rm min}=10^{-9}$ $M_{\odot}$ and for all three annihilation channels the constraints from the extragalactic $\gamma$-ray background completely exclude the PFH favored region. This is also the case for the power law $c(M)$ model with $M_{\rm min}=10^{-6}$ $M_{\odot}$, which for clarity is not shown in the figure. The WMAP $\tau$ constraint, which is insensitive to the structure boost, and thus does not depend on large uncertainties related to $c(M)$, is able to rule out the PFH favored region of \citet{2009arXiv0905.0480M} in the case of $\tau^{-}+\tau^{+}$ channel and also by a large part in the case of $\mu^{-}+\mu^{+}$ annihilation channel. The excluded part of the \citet{2009arXiv0905.0480M} favored region in the $e^{-}+e^{+}$ case is somewhat smaller.

In addition to the extragalactic constraints in Fig. \ref{fig6} we have also shown two types of Galactic bounds as found in \citet{2009arXiv0905.0480M}: (i) constraints derived from the HESS observations of the Galactic Center $\gamma$-rays ($0.1^{\circ}$ angular region), plotted by dashed lines and labeled as GC-$\gamma$, (ii) constraints from the FERMI observations of the $\gamma$-rays below $10$ GeV in the $10^{\circ}<|b|<20^{\circ}$ region (i.e., away from the Galactic Center), plotted by dotted lines and labeled as G-IC$\gamma$.\footnote{At those lower energies the Inverse Compton part of the annihilation $\gamma$-spectrum dominates.} Although both of those Galactic bounds shown here assumed NFW density profile, the G-IC$\gamma$ bound is less dependent on this assumption, as the observed region is away from the Galactic Center. It is evident that in several occasions the complementary extragalactic constraints derived in this work provide stronger bounds on annihilation cross-section, and thus are certainly of considerable interest.

As a final point, note that our constraints based on diffuse extragalactic $\gamma$-ray background depend significantly on the assumed fraction of unresolved astrophysical sources contributing to the EGRET signal. There is a general consensus that a significant fraction of the diffuse signal is due to unresolved AGNs, however the predicted values are highly model dependent with the fractions ranging from $25\%$ up to $100\%$. The upcoming measurements by Fermi satellite should clarify this situation considerably, since compared to EGRET, Fermi has significantly better sensitivity and angular resolution, allowing it to resolve many more individual astrophysical sources \citep{2009ApJ...697.1071A}.

\section{Summary}
In this paper we have constrained observationally motivated leptonic DM annihilation scenario of \citet{2009NuPhB.813....1C} using measurements of the extragalactic $\gamma$-ray background by EGRET space experiment \citep{1998ApJ...494..523S} and the measurement of the optical depth to decoupling by WMAP satellite \citep{2009ApJS..180..306D}. Our main results for all considered leptonic annihilation channels are given on three panels of Fig. \ref{fig6} where we have also added the region favored by the PFH data as given in \citet{2009arXiv0905.0480M}.

Here are our main conclusions:
\begin{itemize}
\item For the power law concentration model $c\propto M^{-0.1}$ the constraint from the extragalactic gamma background rules out the PFH favored region of \citet{2009arXiv0905.0480M} for all three annihilation channels considered in this work.
\item The same constraint in the case of \citet{2008MNRAS.391.1940M} concentration model is significantly weaker owing to the much smaller structure formation boost as seen on the upper panel of Fig. \ref{fig2}. Compared to the mass of the smallest DM halos assumed in this work numerical simulations probe several orders of magnitude larger mass scales. The particular way one chooses to extrapolate the $c(M)$ relation to smaller scales is currently arguably the biggest source of uncertainty for the annihilation-sourced $\gamma$-ray background calculations.
\item For the minimal DM halo masses assumed in this paper the reionization constraints, on the other hand, turn out to be insensitive to the low redshift clustering boost, and thus are free of this large uncertainty in the mass-concentration relation. Thus these constraints are significantly more robust.
\item The primary photons give important contribution to the derived constraints only in the case of $\tau^{-}+\tau^{+}$ annihilation channel.
In the case of  $\mu^{-}+\mu^{+}$ and $e^{-}+e^{+}$ annihilation channels the constraints are mostly given by the IC scattering 
photon flux.
\item For the $\tau^{-}+\tau^{+}$ annihilation channel the constraints from reionization completely rule out the PFH favored region whereas for the $\mu^{-}+\mu^{+}$ case large part of the region gets excluded. In the case of $e^{-}+e^{+}$ annihilation channel the excluded part is relatively small, but one has to keep in mind that in reality this channel does not provide a particularly good fit to the PFH data \citep{2009arXiv0905.0480M}. 
\item In agreement to \citet{2009arXiv0904.1210B} we find that, indeed, it is possible to have annihilating DM models that completely ionize the Universe at low redshifts. However, we find that those models result in a too high ionization fraction plateau at $z\gtrsim 100$, leading to the optical depths which are incompatible with the WMAP measurements. Also, the annihilation cross-section one needs is somewhat larger than suggested by the PFH data.
\item In the light of our results, the foreseen new diffuse $\gamma$-ray data by Fermi satellite  is expected to further constrain 
the DM annihilation solutions to the cosmic ray anomalies.
\end{itemize}

\acknowledgements{We thank  Alessandro Strumia, Jens Chluba and Mario Kadastik for discussions and our Referee for useful comments and suggestions. This work was supported by the ESF Grants 8090, 8005, 7146 and by EU  FP7-INFRA-2007-1.2.3 contract No 223807. GH acknowledges the support provided through a PPARC/STFC postdoctoral fellowship at UCL.}


\begin{thebibliography}{}

\bibitem[{{Abdo} {et~al.}(2009){Abdo}, {Ackermann}, {Ajello}, {Atwood},
  {Axelsson}, {Baldini}, {Ballet}, {Barbiellini}, {Bastieri}, {Battelino},
  {Baughman}, {Bechtol}, {Bellazzini}, {Berenji}, {Blandford}, {Bloom},
  {Bogaert}, {Bonamente}, {Borgland}, {Bregeon}, {Brez}, {Brigida}, {Bruel},
  {Burnett}, {Caliandro}, {Cameron}, {Caraveo}, {Carlson}, {Casandjian},
  {Cecchi}, {Charles}, {Chekhtman}, {Cheung}, {Chiang}, {Ciprini}, {Claus},
  {Cohen-Tanugi}, {Cominsky}, {Conrad}, {Cutini}, {Dermer}, {de Angelis}, {de
  Palma}, {Digel}, {di Bernardo}, {Do Couto E Silva}, {Drell}, {Dubois},
  {Dumora}, {Edmonds}, {Farnier}, {Favuzzi}, {Focke}, {Frailis}, {Fukazawa},
  {Funk}, {Fusco}, {Gaggero}, {Gargano}, {Gasparrini}, {Gehrels}, {Germani},
  {Giebels}, {Giglietto}, {Giordano}, {Glanzman}, {Godfrey}, {Grasso},
  {Grenier}, {Grondin}, {Grove}, {Guillemot}, {Guiriec}, {Hanabata}, {Harding},
  {Hartman}, {Hayashida}, {Hays}, {Hughes}, {J{\'o}hannesson}, {Johnson},
  {Johnson}, {Johnson}, {Kamae}, {Katagiri}, {Kataoka}, {Kawai}, {Kerr},
  {Kn{\"o}dlseder}, {Kocevski}, {Kuehn}, {Kuss}, {Lande}, {Latronico},
  {Lemoine-Goumard}, {Longo}, {Loparco}, {Lott}, {Lovellette}, {Lubrano},
  {Madejski}, {Makeev}, {Massai}, {Mazziotta}, {McConville}, {McEnery},
  {Meurer}, {Michelson}, {Mitthumsiri}, {Mizuno}, {Moiseev}, {Monte},
  {Monzani}, {Moretti}, {Morselli}, {Moskalenko}, {Murgia}, {Nolan}, {Norris},
  {Nuss}, {Ohsugi}, {Omodei}, {Orlando}, {Ormes}, {Ozaki}, {Paneque},
  {Panetta}, {Parent}, {Pelassa}, {Pepe}, {Pesce-Rollins}, {Piron}, {Pohl},
  {Porter}, {Profumo}, {Rain{\`o}}, {Rando}, {Razzano}, {Reimer}, {Reimer},
  {Reposeur}, {Ritz}, {Rochester}, {Rodriguez}, {Romani}, {Roth}, {Ryde},
  {Sadrozinski}, {Sanchez}, {Sander}, {Saz Parkinson}, {Scargle}, {Schalk},
  {Sellerholm}, {Sgr{\`o}}, {Smith}, {Smith}, {Spandre}, {Spinelli}, {Starck},
  {Stephens}, {Strickman}, {Strong}, {Suson}, {Tajima}, {Takahashi},
  {Takahashi}, {Tanaka}, {Thayer}, {Thayer}, {Thompson}, {Tibaldo}, {Tibolla},
  {Torres}, {Tosti}, {Tramacere}, {Uchiyama}, {Usher}, {van Etten},
  {Vasileiou}, {Vilchez}, {Vitale}, {Waite}, {Wallace}, {Wang}, {Winer},
  {Wood}, {Ylinen}, \& {Ziegler}}]{2009PhRvL.102r1101A}
{Abdo}, A.~A., {Ackermann}, M., {Ajello}, M., {et~al.} 2009, Physical Review
  Letters, 102, 181101

\bibitem[{{Adriani} {et~al.}(2009){Adriani}, {Barbarino}, {Bazilevskaya},
  {Bellotti}, {Boezio}, {Bogomolov}, {Bonechi}, {Bongi}, {Bonvicini}, {Bottai},
  {Bruno}, {Cafagna}, {Campana}, {Carlson}, {Casolino}, {Castellini}, {de
  Pascale}, {de Rosa}, {de Simone}, {di Felice}, {Galper}, {Grishantseva},
  {Hofverberg}, {Koldashov}, {Krutkov}, {Kvashnin}, {Leonov}, {Malvezzi},
  {Marcelli}, {Menn}, {Mikhailov}, {Mocchiutti}, {Orsi}, {Osteria}, {Papini},
  {Pearce}, {Picozza}, {Ricci}, {Ricciarini}, {Simon}, {Sparvoli},
  {Spillantini}, {Stozhkov}, {Vacchi}, {Vannuccini}, {Vasilyev}, {Voronov},
  {Yurkin}, {Zampa}, {Zampa}, \& {Zverev}}]{2009Natur.458..607A}
{Adriani}, O., {Barbarino}, G.~C., {Bazilevskaya}, G.~A., {et~al.} 2009, \nat,
  458, 607

\bibitem[{Adriani {et~al.}(2009)}]{Adriani:2008zq}
Adriani, O. {et~al.} 2009, Phys. Rev. Lett., 102, 051101

\bibitem[{Arkani-Hamed {et~al.}(2009)Arkani-Hamed, Finkbeiner, Slatyer, \&
  Weiner}]{ArkaniHamed:2008qn}
Arkani-Hamed, N., Finkbeiner, D.~P., Slatyer, T.~R., \& Weiner, N. 2009, Phys.
  Rev., D79, 015014

\bibitem[{{Atwood} {et~al.}(2009){Atwood}, {Abdo}, {Ackermann}, {Althouse},
  {Anderson}, {Axelsson}, {Baldini}, {Ballet}, {Band}, {Barbiellini},
  {Bartelt}, {Bastieri}, {Baughman}, {Bechtol}, {B{\'e}d{\'e}r{\`e}de},
  {Bellardi}, {Bellazzini}, {Berenji}, {Bignami}, {Bisello}, {Bissaldi},
  {Blandford}, {Bloom}, {Bogart}, {Bonamente}, {Bonnell}, {Borgland},
  {Bouvier}, {Bregeon}, {Brez}, {Brigida}, {Bruel}, {Burnett}, {Busetto},
  {Caliandro}, {Cameron}, {Caraveo}, {Carius}, {Carlson}, {Casandjian},
  {Cavazzuti}, {Ceccanti}, {Cecchi}, {Charles}, {Chekhtman}, {Cheung},
  {Chiang}, {Chipaux}, {Cillis}, {Ciprini}, {Claus}, {Cohen-Tanugi},
  {Condamoor}, {Conrad}, {Corbet}, {Corucci}, {Costamante}, {Cutini}, {Davis},
  {Decotigny}, {DeKlotz}, {Dermer}, {de Angelis}, {Digel}, {do Couto e Silva},
  {Drell}, {Dubois}, {Dumora}, {Edmonds}, {Fabiani}, {Farnier}, {Favuzzi},
  {Flath}, {Fleury}, {Focke}, {Funk}, {Fusco}, {Gargano}, {Gasparrini},
  {Gehrels}, {Gentit}, {Germani}, {Giebels}, {Giglietto}, {Giommi}, {Giordano},
  {Glanzman}, {Godfrey}, {Grenier}, {Grondin}, {Grove}, {Guillemot}, {Guiriec},
  {Haller}, {Harding}, {Hart}, {Hays}, {Healey}, {Hirayama}, {Hjalmarsdotter},
  {Horn}, {Hughes}, {J{\'o}hannesson}, {Johansson}, {Johnson}, {Johnson},
  {Johnson}, {Johnson}, {Kamae}, {Katagiri}, {Kataoka}, {Kavelaars}, {Kawai},
  {Kelly}, {Kerr}, {Klamra}, {Kn{\"o}dlseder}, {Kocian}, {Komin}, {Kuehn},
  {Kuss}, {Landriu}, {Latronico}, {Lee}, {Lee}, {Lemoine-Goumard}, {Lionetto},
  {Longo}, {Loparco}, {Lott}, {Lovellette}, {Lubrano}, {Madejski}, {Makeev},
  {Marangelli}, {Massai}, {Mazziotta}, {McEnery}, {Menon}, {Meurer},
  {Michelson}, {Minuti}, {Mirizzi}, {Mitthumsiri}, {Mizuno}, {Moiseev},
  {Monte}, {Monzani}, {Moretti}, {Morselli}, {Moskalenko}, {Murgia},
  {Nakamori}, {Nishino}, {Nolan}, {Norris}, {Nuss}, {Ohno}, {Ohsugi}, {Omodei},
  {Orlando}, {Ormes}, {Paccagnella}, {Paneque}, {Panetta}, {Parent}, {Pearce},
  {Pepe}, {Perazzo}, {Pesce-Rollins}, {Picozza}, {Pieri}, {Pinchera}, {Piron},
  {Porter}, {Poupard}, {Rain{\`o}}, {Rando}, {Rapposelli}, {Razzano}, {Reimer},
  {Reimer}, {Reposeur}, {Reyes}, {Ritz}, {Rochester}, {Rodriguez}, {Romani},
  {Roth}, {Russell}, {Ryde}, {Sabatini}, {Sadrozinski}, {Sanchez}, {Sander},
  {Sapozhnikov}, {Parkinson}, {Scargle}, {Schalk}, {Scolieri}, {Sgr{\`o}},
  {Share}, {Shaw}, {Shimokawabe}, {Shrader}, {Sierpowska-Bartosik}, {Siskind},
  {Smith}, {Smith}, {Spandre}, {Spinelli}, {Starck}, {Stephens}, {Strickman},
  {Strong}, {Suson}, {Tajima}, {Takahashi}, {Takahashi}, {Tanaka}, {Tenze},
  {Tether}, {Thayer}, {Thayer}, {Thompson}, {Tibaldo}, {Tibolla}, {Torres},
  {Tosti}, {Tramacere}, {Turri}, {Usher}, {Vilchez}, {Vitale}, {Wang},
  {Watters}, {Winer}, {Wood}, {Ylinen}, \& {Ziegler}}]{2009ApJ...697.1071A}
{Atwood}, W.~B., {Abdo}, A.~A., {Ackermann}, M., {et~al.} 2009, \apj, 697, 1071

\bibitem[{{Belikov} \& {Hooper}(2009{\natexlab{a}})}]{2009arXiv0904.1210B}
{Belikov}, A.~V. \& {Hooper}, D. 2009{\natexlab{a}}, ArXiv 0904.1210

\bibitem[{{Belikov} \& {Hooper}(2009{\natexlab{b}})}]{2009arXiv0906.2251B}
{Belikov}, A.~V. \& {Hooper}, D. 2009{\natexlab{b}}, ArXiv 0906.2251

\bibitem[{{Bell} \& {Jacques}(2009)}]{2009PhRvD..79d3507B}
{Bell}, N.~F. \& {Jacques}, T.~D. 2009, \prd, 79, 043507

\bibitem[{Bertone {et~al.}(2009)Bertone, Cirelli, Strumia, \&
  Taoso}]{Bertone:2008xr}
Bertone, G., Cirelli, M., Strumia, A., \& Taoso, M. 2009, JCAP, 0903, 009

\bibitem[{{Borriello} {et~al.}(2009){Borriello}, {Cuoco}, \&
  {Miele}}]{2009arXiv0903.1852B}
{Borriello}, E., {Cuoco}, A., \& {Miele}, G. 2009, ArXiv 0903.1852

\bibitem[{{Bringmann}(2009)}]{2009arXiv0903.0189B}
{Bringmann}, T. 2009, ArXiv 0903.0189

\bibitem[{{Chang} {et~al.}(2008){Chang}, {Adams}, {Ahn}, {Bashindzhagyan},
  {Christl}, {Ganel}, {Guzik}, {Isbert}, {Kim}, {Kuznetsov}, {Panasyuk},
  {Panov}, {Schmidt}, {Seo}, {Sokolskaya}, {Watts}, {Wefel}, {Wu}, \&
  {Zatsepin}}]{2008Natur.456..362C}
{Chang}, J., {Adams}, J.~H., {Ahn}, H.~S., {et~al.} 2008, \nat, 456, 362

\bibitem[{{Chen} \& {Kamionkowski}(2004)}]{2004PhRvD..70d3502C}
{Chen}, X. \& {Kamionkowski}, M. 2004, \prd, 70, 043502

\bibitem[{{Chiang} \& {Mukherjee}(1998)}]{1998ApJ...496..752C}
{Chiang}, J. \& {Mukherjee}, R. 1998, \apj, 496, 752

\bibitem[{{Cirelli} {et~al.}(2009){Cirelli}, {Kadastik}, {Raidal}, \&
  {Strumia}}]{2009NuPhB.813....1C}
{Cirelli}, M., {Kadastik}, M., {Raidal}, M., \& {Strumia}, A. 2009, Nuclear
  Physics B, 813, 1

\bibitem[{{Cirelli} \& {Panci}(2009)}]{2009arXiv0904.3830C}
{Cirelli}, M. \& {Panci}, P. 2009, ArXiv 0904.3830

\bibitem[{{Cooray} \& {Sheth}(2002)}]{2002PhR...372....1C}
{Cooray}, A. \& {Sheth}, R. 2002, \physrep, 372, 1

\bibitem[{{Dunkley} {et~al.}(2009){Dunkley}, {Komatsu}, {Nolta}, {Spergel},
  {Larson}, {Hinshaw}, {Page}, {Bennett}, {Gold}, {Jarosik}, {Weiland},
  {Halpern}, {Hill}, {Kogut}, {Limon}, {Meyer}, {Tucker}, {Wollack}, \&
  {Wright}}]{2009ApJS..180..306D}
{Dunkley}, J., {Komatsu}, E., {Nolta}, M.~R., {et~al.} 2009, \apjs, 180, 306

\bibitem[{{Fan} {et~al.}(2006){Fan}, {Carilli}, \&
  {Keating}}]{2006ARA&A..44..415F}
{Fan}, X., {Carilli}, C.~L., \& {Keating}, B. 2006, \araa, 44, 415

\bibitem[{{Galli} {et~al.}(2009){Galli}, {Iocco}, {Bertone}, \&
  {Melchiorri}}]{2009arXiv0905.0003G}
{Galli}, S., {Iocco}, F., {Bertone}, G., \& {Melchiorri}, A. 2009, ArXiv
  0905.0003

\bibitem[{{HESS Collaboration: F.~Aharonian}(2009)}]{2009arXiv0906.2002H}
{HESS Collaboration: F.~Aharonian}. 2009, ArXiv 0906.2002

\bibitem[{{Kawasaki} {et~al.}(2009){Kawasaki}, {Kohri}, \&
  {Nakayama}}]{2009arXiv0904.3626K}
{Kawasaki}, M., {Kohri}, K., \& {Nakayama}, K. 2009, ArXiv 0904.3626

\bibitem[{{Macci{\`o}} {et~al.}(2008){Macci{\`o}}, {Dutton}, \& {van den
  Bosch}}]{2008MNRAS.391.1940M}
{Macci{\`o}}, A.~V., {Dutton}, A.~A., \& {van den Bosch}, F.~C. 2008, \mnras,
  391, 1940

\bibitem[{{Mapelli} {et~al.}(2006){Mapelli}, {Ferrara}, \&
  {Pierpaoli}}]{2006MNRAS.369.1719M}
{Mapelli}, M., {Ferrara}, A., \& {Pierpaoli}, E. 2006, \mnras, 369, 1719

\bibitem[{{Martinez} {et~al.}(2009){Martinez}, {Bullock}, {Kaplinghat},
  {Strigari}, \& {Trotta}}]{2009JCAP...06..014M}
{Martinez}, G.~D., {Bullock}, J.~S., {Kaplinghat}, M., {Strigari}, L.~E., \&
  {Trotta}, R. 2009, Journal of Cosmology and Astro-Particle Physics, 6, 14

\bibitem[{{Meade} {et~al.}(2009){Meade}, {Papucci}, {Strumia}, \&
  {Volansky}}]{2009arXiv0905.0480M}
{Meade}, P., {Papucci}, M., {Strumia}, A., \& {Volansky}, T. 2009, ArXiv
  0905.0480

\bibitem[{{M{\"u}cke} \& {Pohl}(2000)}]{2000MNRAS.312..177M}
{M{\"u}cke}, A. \& {Pohl}, M. 2000, \mnras, 312, 177

\bibitem[{{Mukherjee} \& {Chiang}(1999)}]{1999APh....11..213M}
{Mukherjee}, R. \& {Chiang}, J. 1999, Astroparticle Physics, 11, 213

\bibitem[{{Natarajan} \& {Schwarz}(2008)}]{2008PhRvD..78j3524N}
{Natarajan}, A. \& {Schwarz}, D.~J. 2008, \prd, 78, 103524

\bibitem[{{Navarro} {et~al.}(1997){Navarro}, {Frenk}, \&
  {White}}]{1997ApJ...490..493N}
{Navarro}, J.~F., {Frenk}, C.~S., \& {White}, S.~D.~M. 1997, \apj, 490, 493

\bibitem[{{Padmanabhan} \& {Finkbeiner}(2005)}]{2005PhRvD..72b3508P}
{Padmanabhan}, N. \& {Finkbeiner}, D.~P. 2005, \prd, 72, 023508

\bibitem[{{Profumo} \& {Jeltema}(2009)}]{2009arXiv0906.0001P}
{Profumo}, S. \& {Jeltema}, T.~E. 2009, ArXiv 0906.0001

\bibitem[{{Seager} {et~al.}(1999){Seager}, {Sasselov}, \&
  {Scott}}]{1999ApJ...523L...1S}
{Seager}, S., {Sasselov}, D.~D., \& {Scott}, D. 1999, \apjl, 523, L1

\bibitem[{{Sheth} \& {Tormen}(1999)}]{1999MNRAS.308..119S}
{Sheth}, R.~K. \& {Tormen}, G. 1999, \mnras, 308, 119

\bibitem[{{Shull} \& {van Steenberg}(1985)}]{1985ApJ...298..268S}
{Shull}, J.~M. \& {van Steenberg}, M.~E. 1985, \apj, 298, 268

\bibitem[{{Sj{\"o}strand} {et~al.}(2008){Sj{\"o}strand}, {Mrenna}, \&
  {Skands}}]{2008CoPhC.178..852S}
{Sj{\"o}strand}, T., {Mrenna}, S., \& {Skands}, P. 2008, Computer Physics
  Communications, 178, 852

\bibitem[{{Slatyer} {et~al.}(2009){Slatyer}, {Padmanabhan}, \&
  {Finkbeiner}}]{2009arXiv0906.1197S}
{Slatyer}, T.~R., {Padmanabhan}, N., \& {Finkbeiner}, D.~P. 2009, ArXiv
  0906.1197

\bibitem[{{Sreekumar} {et~al.}(1998){Sreekumar}, {Bertsch}, {Dingus},
  {Esposito}, {Fichtel}, {Hartman}, {Hunter}, {Kanbach}, {Kniffen}, {Lin},
  {Mayer-Hasselwander}, {Michelson}, {von Montigny}, {Muecke}, {Mukherjee},
  {Nolan}, {Pohl}, {Reimer}, {Schneid}, {Stacy}, {Stecker}, {Thompson}, \&
  {Willis}}]{1998ApJ...494..523S}
{Sreekumar}, P., {Bertsch}, D.~L., {Dingus}, B.~L., {et~al.} 1998, \apj, 494,
  523

\bibitem[{{Stecker} \& {Salamon}(1996)}]{1996ApJ...464..600S}
{Stecker}, F.~W. \& {Salamon}, M.~H. 1996, \apj, 464, 600

\bibitem[{{Strong} {et~al.}(2004){Strong}, {Moskalenko}, \&
  {Reimer}}]{2004ApJ...613..956S}
{Strong}, A.~W., {Moskalenko}, I.~V., \& {Reimer}, O. 2004, \apj, 613, 956

\bibitem[{{Torii} {et~al.}(2008){Torii}, {Yamagami}, {Tamura}, {Yoshida},
  {Kitamura}, {Anraku}, {Chang}, {Ejiri}, {Iijima}, {Kadokura}, {Kasahara},
  {Katayose}, {Kobayashi}, {Komori}, {Matsuzaka}, {Mizutani}, {Murakami},
  {Namiki}, {Nishimura}, {Ohta}, {Saito}, {Shibata}, {Tateyama}, {Yamagishi},
  {Yamashita}, \& {Yuda}}]{2008arXiv0809.0760T}
{Torii}, S., {Yamagami}, T., {Tamura}, T., {et~al.} 2008, ArXiv 0809.0760

\bibitem[{{Zdziarski} \& {Svensson}(1989)}]{1989ApJ...344..551Z}
{Zdziarski}, A.~A. \& {Svensson}, R. 1989, \apj, 344, 551

\bibitem[{{Zhang} {et~al.}(2006){Zhang}, {Chen}, {Lei}, \&
  {Si}}]{2006PhRvD..74j3519Z}
{Zhang}, L., {Chen}, X., {Lei}, Y.-A., \& {Si}, Z.-G. 2006, \prd, 74, 103519

\end{thebibliography}
\end{document}